\documentclass{elsart}
\usepackage{graphicx}
\usepackage{graphics}
\usepackage{amssymb}
\usepackage{amsmath}
\usepackage{bm}
\usepackage{lineno}

\newcommand{\be}{\begin{equation}}
\newcommand{\ee}{\end{equation}}
\newcommand{\bea}{\begin{eqnarray}}
\newcommand{\eea}{\end{eqnarray}}
\newcommand{\beal}{\begin{align}}
\newcommand{\eal}{\end{align}}
\newcommand{\bespl}{\begin{split}}
\newcommand{\espl}{\end{split}}

\newcommand{\nslash}{\kern 0.2 em n\kern -0.50em /}
\newcommand{\kslash}{\kern 0.2 em k\kern -0.45em /}
\newcommand{\pslash}{\kern 0.2 em p\kern -0.50em /}
\newcommand{\Sslash}{\kern 0.2 em S\kern -0.50em /}
\newcommand{\Pslash}{\kern 0.2 em P\kern -0.50em /}
\newcommand{\Rslash}{\kern 0.2 em R\kern -0.50em /}
\begin{document}
%\linenumbers

\begin{frontmatter}

\title{At the borderline between exclusive and inclusive physics:
Study of Drell-Yan fragments in the PANDA experiment (A preliminary 
simulation). 
} 

\author{A.~Bianconi}
\address{Dipartimento di Chimica e Fisica per l'Ingegneria e per i 
Materiali, Universit\`a di Brescia, I-25123 Brescia, Italy, and\\
Istituto Nazionale di Fisica Nucleare, Sezione di Pavia, I-27100 Pavia, 
Italy}
\ead{andrea.bianconi@bs.infn.it}

%%%%%%%%%%%%%%%%%%%%%%%%%%%%%%%%%%%%%%%%%%%%%%%%%%%%%%%%%%%%
\begin{abstract}
Here a preliminary study is presented concerning the detection 
of the normally unseen 
Drell-Yan fragments, possible in the PANDA experiment. To work 
as a multi-purpose apparatus, this 
experiment will 
record all the particles produced in the collisions between the 
antiproton beam and the target, with a rather wide acceptance. 
So detecting Drell-Yan dileptons 
with or without analyzing the other fragments is just a matter of 
applying cutoffs in the data analysis stage. 
The distribution of 
the products of 50,000 typical Drell-Yan 
events is here simulateded using 
a well-known generator code (Pythia-8). The resulting distributions 
are inserted within the PANDA acceptance region, to analyze 
the chances of missing some searched fragment combinations, or of 
confusing different sets of particles. 
The most interesting result is that, due to 
the reduced phase space, the produced states 
are much simpler than one  could imagine: (i) 
almost 50 \% of the 
events just consist of a dilepton plus a nucleon-antinucleon pair; 
(ii) practically all events present a nucleon-antinucleon pair; 
(iii) the number of light particles (photons over an infrared cutoff 
and pions) is pretty small. The presented simulations 
show that it is possible to study 
experimentally some, or some aspects, of the most relevant final states,   
with good statistics and precision. 
\end{abstract} 


\end{frontmatter}

\maketitle

\section{Introduction}

The PANDA experiment\cite{panda} has a rather wide program, that 
includes electromagnetic and strong hard processes 
in collisions between antiproton beams and 
unpolarized nuclear targets (see ref.\cite{PandaPhysicsBook}, 
or more synthetically ref.\cite{Bettoni09}). 
The center-of-mass energy will range from small values up 
to about $\sqrt{30}$ GeV (beam energy ranging from 1 to 13 GeV).  
Among the planned ones, 
measurements of Drell-Yan processes\cite{DrellYan} 
will be performed in the 
highest part of the beam energy spectrum, with attention to 
both di-muon and di-electron production. 

Drell-Yan is a very well studied class of inclusive processes, 
however up to date very little information is available on the 
(normally undetected) fragments produced in association with 
the Drell-Yan lepton pair. What is available 
(essentially: charge multiplicity)
was measured in 
collider experiments at very large center of mass 
energy (e.g.\cite{Antreasyan88}, 
where $\sqrt{s}$ $\sim$ 50-100 GeV). 

No information of this kind is available from fixed-target 
experiments at much lower $\sqrt{s}$. 
The point is that the really appealing aspect of a Drell-Yan 
measurement is that one needs $not$ to care these particles, since  
one may build a relatively simple apparatus where a thick screen 
absorbes everything but muons, and muons are analyzed downstream 
with respect to the screens. So, only multi-purpose apparata 
may be suitable to analyze hadronic or electromagnetic 
fragments that accompany the dilepton production.

PANDA is a multipurpose apparatus, with a 
very peculiar feature: its on-line filters will be reduced to a minimum, 
so that at each collision 
all 
the following fragments  
will be detected (within the large experiment acceptance). 
The idea is to record everything, to later select 
different processes in the data analysis phase. 

To understand what kind of phenomenology one could meet, 
I have performed some simulations, using the popular high-energy 
Montecarlo generator 
Pythia-8 \cite{Pythia8}. 
Although this is $not$ the 
energy regime for which this code has been optimized, 
some features 
determining the fragments are so universal (in particular in the 
PANDA case: the 
size of the available phase space) that the gross features of the 
predicted distributions should be reliable. 
\footnote{The 
author of this work is the author of a 
Drell-Yan generator code\cite{DY_AB5} that has been extensively 
used for preliminary studies of PANDA dilepton 
distributions\cite{BR06a,AB06,BR_JPG2}. 
Although this code is optimized on the conditions of the 
PANDA case, it does not produce fragments, apart for the 
lepton-antilepton pair. 

It must be signalled that 
the default parameters of Pythia 
include some phase space constraints 
that must be modified to access the 
PANDA kinematical region. 
} 
And these features 
leave room for interesting 
perspectives.

Here I have studied the class of Drell-Yan events where the 
dilepton mass and transverse momentum are $>$ 2 GeV/c$^2$ and 
$>$ 0.8 GeV/c respectively. 
Below the mass 1.8 GeV/c$^2$, the effect of the tails of the 
vector resonances is strong, 
and it may be ambiguous to associate dileptons to Drell-Yan 
events. 
We must remark that at 
increasing masses the event rates become increasingly suppressed, 
roughly as $1/M^2$. So, most of the Drell-Yan events 
considered here 
have a dilepton mass near 2 GeV/c$^2$, and this will be 
the situation in PANDA (see refs. 
\cite{BR06a,AB06,BR_JPG2} for systematic studies 
of the expected dilepton distributions in PANDA). 
In present-day unpolarized Drell-Yan, the 
most interesting observables 
are those associated with azimuthal 
asymmetries (\cite{LamTung78,BoerMulders98}, see 
also \cite{ArnoldMetzSchlegel09} 
for a very 
general and systematic treatment of the cross section structure). 
These become 
visible at transverse momenta of magnitude 1-2 
GeV/c (see the previous measurements by 
refs. \cite{NA3,NA10,Conway89}). 
Taking into account that the Drell-Yan event rates will be 
peaked at transverse momenta $\approx$ $0.5-1$ GeV/c, an event with 
$q_T$ $=$ 
1 GeV/c will be the most common among the interesting 
ones\cite{BR06a}. 

Pythia does not include detailed features like azimuthal asymmetries. 
So what is simulated here is $not$ exactly what we will see in the 
experiment, 
where it  will be interesting, and new, to be able to study  
joint azimuthal distributions of the dilepton and of the fragments. 
But the Pythia-based analysis presented here 
may tell us which kind of fragment combinations it  will make sense 
to analyze. 

The most appealing result of the following simulations is 
that Drell-Yan 
is, in PANDA conditions, a ``quasi-exclusive'' process. Potentially this 
will allow for theoretical modelling in a way that would be unusual 
from the point of view of traditional Drell-Yan physics.

\section{Drell-Yan fragments in PANDA - simulations}

\subsection{General features of the simulation}

50,000 Drell-Yan events have been here sorted using Pythia-8, 
according to what 
has been imagined as a reasonable kinematical setup for PANDA: 
Antiproton beam energy 15 GeV, at-rest target protons, dilepton 
minimum mass 2 GeV/c$^2$ and dilepton minimum transverse momentum 
0.8 GeV/c.

\begin{figure}[ht]
\centering
\includegraphics[width=9cm]{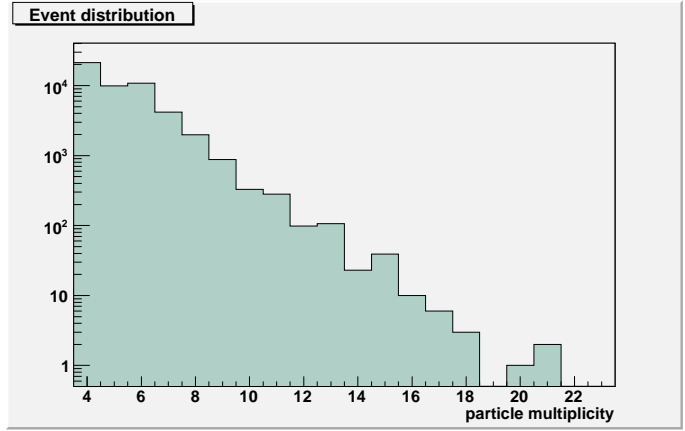}
\caption{Multiplicity of final particles in Drell-Yan dilepton 
production, including the lepton pair (so, by definition $N$ $>$ 2
here). 
The cutoffs on the dilepton mass and transverse momentum are: 
$M$ $>$ 2 GeV/c$^2$, $q_T$ $>$ 0.8 GeV/c. 
}
\end{figure}

For real photons in the final state, I have included a lower 
cutoff on the detection: $E_\gamma$ $>$ 0.2 GeV. The number 
of radiated photons with arbitrarily low energy is in principle 
infinite. So the particle multiplicities in the following must 
be understood as excluding ``soft'' photons. 

Neutral pions do not appear directly in the final state. They 
are present via their offspring photons. So, in the following 
``pions'' means ``charged pions''.

\subsection{Fragment multiplicities}

Figure 1 and 
tables 1, 2 and 3 are devoted to event multiplicities. They 
include the lepton-antilepton pair into the 
total event multiplicity. So, the lowest recorded event multiplicity is 
4 (normally: lepton, antilepton, nucleon, antinucleon).

\begin{quote}
\begin{tabular}{|r|r|}
\hline
total number of events & 50000 \\
\hline
events with no (anti)baryons & 179 \\
\hline
events with 1 $N\bar{N}$ pair  & 49805 \\
\hline
events with 2 $N\bar{N}$ pairs & 16 \\
\hline
events with a $p\bar{p}$ pair & 21765 \\ 
\hline
events with an $n\bar{n}$ pair & 20078 \\ 
\hline
events with a $p\bar{n}$ or $n\bar{p}$ pair & 7993 \\ 
\hline
events with a $p$ & 25761 \\ 
\hline
events with a $\bar{p}$ & 25754 \\ 
\hline
events with an $n$ & 24068 \\
\hline
events with a $\bar{n}$ & 24074 \\
\hline
\end{tabular}
\\ \\ Table 1: Composition of the final state for the simulated 
set of 50,000 events presented here. 
All these events are taken with cutoffs $M$ $>$ 2 GeV/c$^2$, 
$q_T$ $>$ 0.8 GeV/c, on the dilepton kinematics. From the 4th row 
(events with $p\bar{p}$ pair) 
``events with a ...'' means ``events with a ... at least''. 
\end{quote}

\noindent
The most interesting found features are: 

1) 99.7 \% of the events contain a nucleon-antinucleon pair. 

2) Among these, the 
relative numbers of $p\bar{p}$, $n\bar{n}$ and mixed 
($p\bar{n}$ or $n\bar{p}$) pairs are roughly 11:10:4. 

3) 43 \% of the events contain four particles only: a dilepton pair and 
an $N\bar{N}$ pair (same flavor of course). 
The mixed $N\bar{N'}$ 
pairs are normally accompanied by one charged pion. 

4) Special 
interest have the $p\bar{p}-$only events, i.e. events 
where the dilepton is only accompanied by a $p\bar{p}$ pair. 
These are 10,747, i.e. 
49 \% of the total number 21,765 
of events containing a $p\bar{p}$ pair (compare Table 1 and 2), 
and 22 \% of the total number of all events of any kind. 

5) In 42 \% of the events the $N\bar{N}'$ pair is accompanied 
by one/two light particles (mostly 
photons, but frequently charged pions too). 

6) In 15 \% of the events the $N\bar{N}'$ pair is accompanied by 
more than two light particles. 

7) A very small number of events presents no nucleon-antinucleon 
pairs, or two nucleon-antinucleon pairs,  
or a very large light particle multiplicity (15$-$20 hadrons).  

8) The total number of emitted photons is $\approx$ 34,700, the total 
number of charged pions is $\approx$ 15,200, almost equally divided 
into positive and negative ones. Assuming that positive, negative 
and neutral pions are produced with similar rates, half of the photons 
are decay products of a $\pi^o$.

%\newpage

\begin{quote}
\begin{tabular}{|r|r|r|r|r|r|}
\hline
total number & events & events with & events & events & events with \\ 
of final particles & with no & antinucleon- & with a $p\bar{p}$ 
& with a $n\bar{n}$ & a $p\bar{n}$ or  $n\bar{p}$ \\ 
(including $l^+l^-$) & baryon & -nucleon pair & pair  & pair  & pair \\ 
\hline
4 & 1 &   21286 &     10747 &     10538 &     0  \\    
5 & 1 &   9897 &     2793 &     2764 &     4338  \\    
6 & 19 &  10863 &     5428 &     4306 &     1110 \\    
7 & 9 &  4186 &     1521 &     1086 &     1578 \\     
8 & 50 &  1990 &     708 &     703 &     531 \\     
9 & 11 &  880 &     290 &     346 &     234 \\     
10 & 37 &  329 &     97 &     93 &     106 \\     
11 & 15 &  281 &     101 &     136 &     29 \\     
12 & 15 &  98 &     15 &     25 &     43 \\     
13 & 11 &  106 &     40 &     53 &     2 \\     
14 & 2 &  23 &     6 &     3 &     12 \\     
15 & 2 &  39 &     17 &     17 &     3 \\     
$>$15 & 5 &  22 &   2 &   8 &     7 \\     
\hline
\end{tabular}
\\
\\Table 2: Distribution of the events associated with given final particle 
multiplicities and with the presence of (one, two or none) specific 
nucleon-antinucleon pair. The numbers of particles in the left column 
include this 
pair, the $e^+e^-$ or $\mu^+\mu^-$ pair characterizing a Drell-Yan event,
charged pions 
and photons (some coming from neutral pions). 
\end{quote}

In first approximation the equality between 
$\bar{p}p$ and $\bar{n}n$ 
final pairs may be interpreted in terms of a very simple statistical 
model: Assuming pure valence-valence Drell-Yan quark-antiquark 
annihilations, with probability 8/9 the annihilation is $u\bar{u}$, 
leaving a diquark-antidiquark state formed by a 
$ud$ and a $\bar{u}\bar{d}$. 
If we attribute equal probability 50 \% to the vacuum creation 
of a single extra 
$u\bar{u}$ or $d\bar{d}$ pair, we just 
get $\bar{p}p$ and $\bar{n}n$ pairs with the same probability. 

We notice that (i) we still have a probability 1/9 to 
annihilate $d\bar{d}$ and leave a 
diquark-antidiquark state formed by a 
$uu$ and a $\bar{u}\bar{u}$, (ii) the diquarks and the 
new-created quarks may rearrange, with a smaller probability 
because of the small phase space, 
into a Delta, instead of a nucleon state. 

If we decide that a Delta-antiDelta pair 
is just forbidden, than the (produced with 1/9-probability) 
$uu$ and a $\bar{u}\bar{u}$ diquarks will rearrange, in 50 \% 
of the cases, into a $p\bar{p}$ pair, and that is the end of the 
story. So, the overall probability ratio of $p\bar{p}$ 
to $n\bar{n}$ is $4.5:4$ i.e. $9:8$, that is not too far 
from the $11:10$ Pythia-simulated ratio.

\begin{quote}
\begin{tabular}{|r|r|r|}
\hline
 N & events with & events with \\
 & N charged pions & N gammas \\
\hline
0 & 38882 & 31645 \\
1 & 7836 & 7051 \\
2 & 2814 & 8411 \\
3 & 192 & 1678 \\
4 & 225 & 723 \\
5 & 25 & 275 \\
6 & 25 & 99 \\
7 & 1 & 68 \\
8 & 0 & 19 \\
9 & 0 & 18 \\
10 & 0 & 9 \\
$>$10 & 0 & 4 \\
\hline
\end{tabular}
\\
\\Table 3: distribution of the number of events presenting a given 
multiplicity of final charged pions or of final photons. 
Photons are subject to the cutoff $E_\gamma$ $>$ 0.2 GeV. 
Neutral  
pions are ``hidden'' in the photon pairs produced by their decay. 
\end{quote}

\begin{figure}[ht]
\centering
\includegraphics[width=9cm]{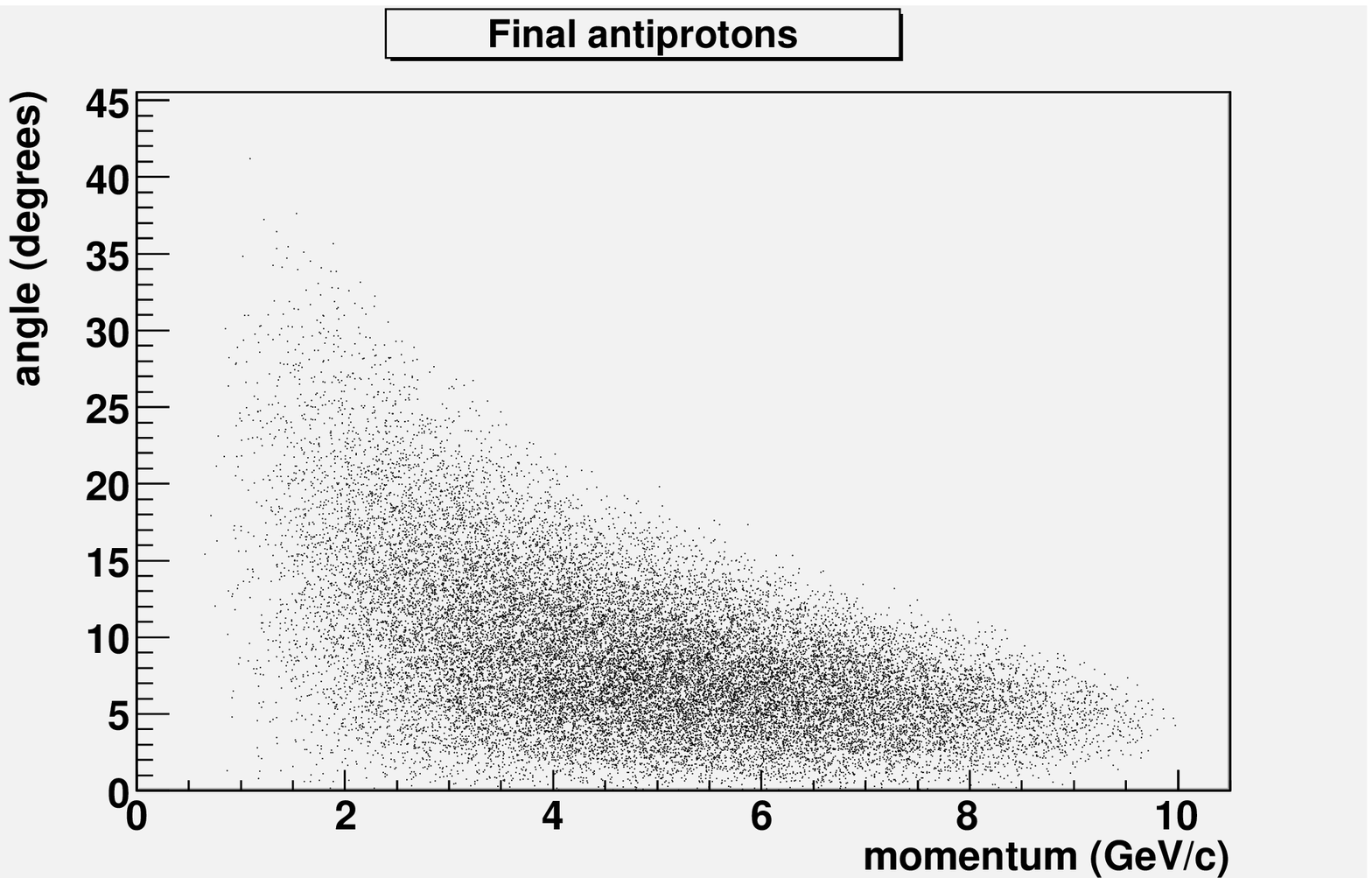}
\caption{
Scatter plot of the main kinematical variables for the 
antiprotons accompanying dilepton production. 
The cutoffs on the dilepton mass and transverse momentum are: 
$M$ $>$ 2 GeV/c$^2$, $q_T$ $>$ 0.8 GeV/c. 
}
\end{figure}

\begin{figure}[ht]
\centering
\includegraphics[width=9cm]{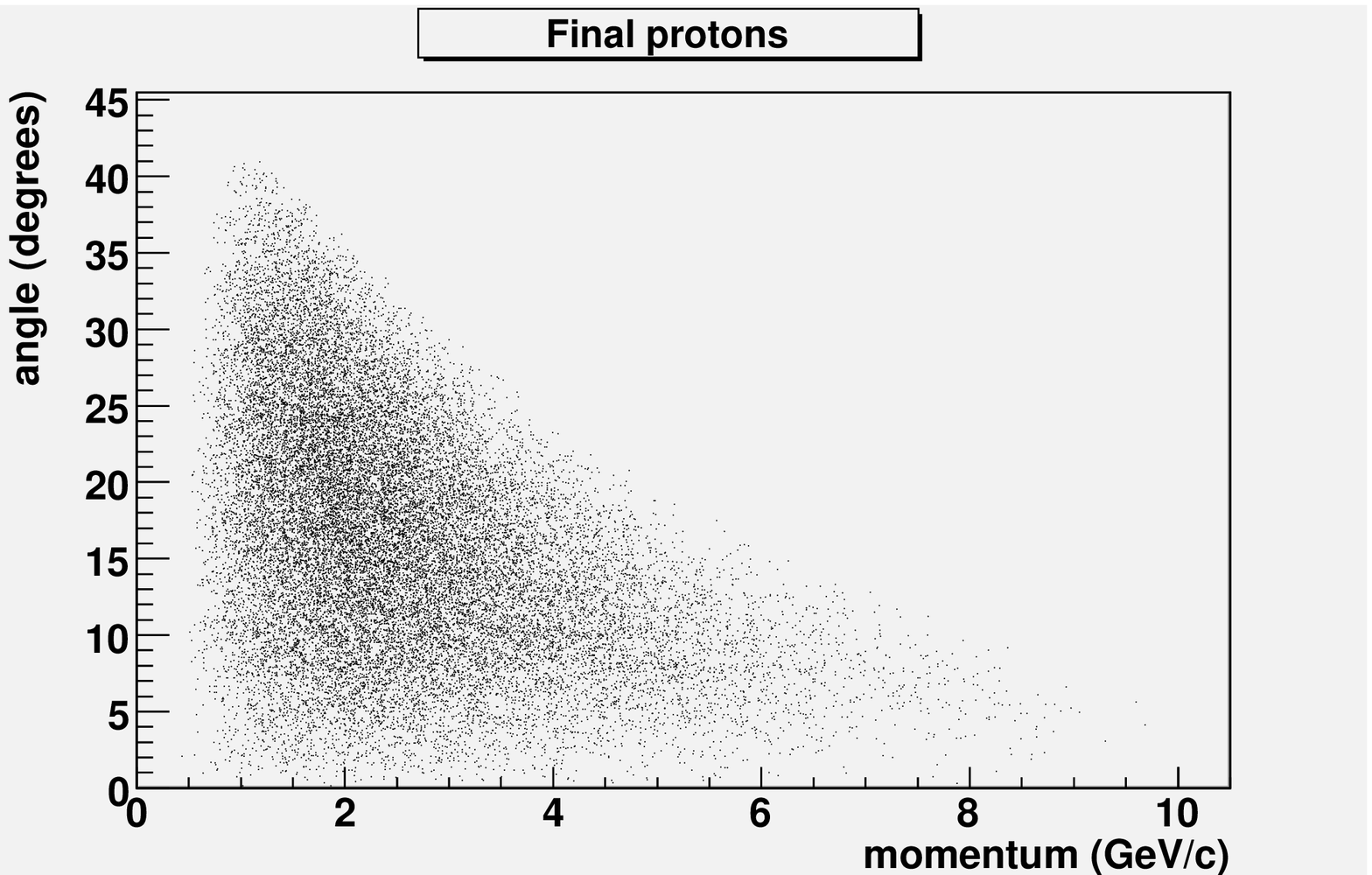}
\caption{
Scatter plot of the main kinematical variables for the 
protons accompanying dilepton production. 
The cutoffs on the dilepton mass and transverse momentum are: 
$M$ $>$ 2 GeV/c$^2$, $q_T$ $>$ 0.8 GeV/c. 
}
\end{figure}

\begin{figure}[ht]
\centering
\includegraphics[width=9cm]{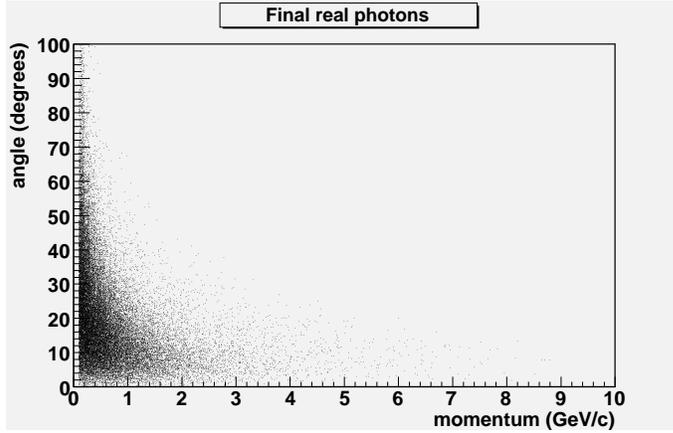}
\caption{
Scatter plot of the main kinematical variables for the 
photons accompanying dilepton production. 
The cutoffs on the dilepton mass and transverse momentum are: 
$M$ $>$ 2 GeV/c$^2$, $q_T$ $>$ 0.8 GeV/c. 
}
\end{figure}

According to Tables 1-2, in almost half of the simulated events 
the fragment production only consisted in the just described 
$q\bar{q}-$creation with simple probabilistic features and 
no further particle production. 
This is very interesting, since once a reasonable 
statistics on $l^+l^-+p\bar{p}$ is accumulated in the experiment, it 
will be possible to examine several observables related to the 
correlations between these particles, and the theoretical 
analysis will be comparatively simple, since these are $exclusive$ 
few-particle events. 

\subsection{Kinematic distributions}

The statistical ``model'' based on $q\bar{q}-$pair 
vacuum-creation plus 
diquark-quark ricombination of the previous subsection suggests a 
strong degree of kinematical continuity between reacting particles 
and final particles with the same identity. This is confirmed 
by figures 2 and 3, where the momentum-angle scatter plot of 
the final antiprotons and protons are reported, and clearly the 
antiproton is much more beam-oriented 
than the proton. 

Protons and antiprotons cover a small angular range. None of the 
simulated protons or antiprotons is produced with angles 
over 45$^o$. 
In the case 
of the lighter particles, we see a much longer angular tail (we 
find rare events also at 170$^o$, although they are not shown 
in the figures), but still the largest part of the light particle 
distribution concentrates at angles below 40$^o$, as in the 
proton and antiproton cases. 
The real difference is in the momentum: far the largest part 
of the pions and 
photons is found below 1 GeV/c, while for protons and 
antiprotons it is the opposite. 

\begin{figure}[ht]
\centering
\includegraphics[width=9cm]{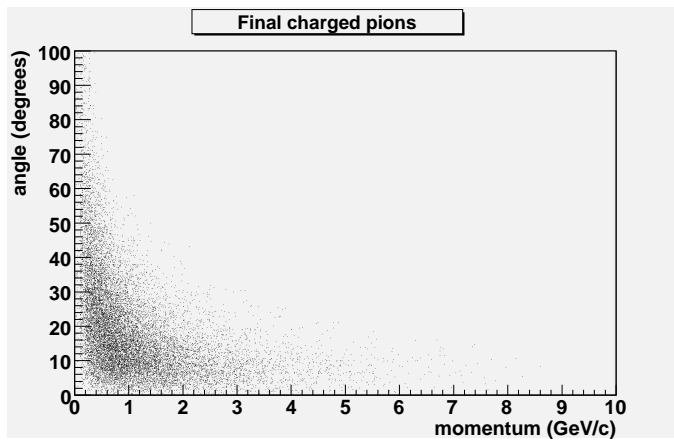}
\caption{
Scatter plot of the main kinematical variables for the 
charged (positive and negative) pions accompanying dilepton production. 
The cutoffs on the dilepton mass and transverse momentum are: 
$M$ $>$ 2 GeV/c$^2$, $q_T$ $>$ 0.8 GeV/c. 
}
\end{figure}

In the case of the pion scatter plot, 
I have chosen to put together negative and positive pions 
after first examining two apparently identical individual scatter plots. 
In Table 4, I report the percentual difference between the population 
of positive and negative pions in the most populated ``macro-bins'' 
(10 degrees times 1 GeV is the range of each 
2-dimensional bin, large enough to allow many bins to 
reach a population of 200 particles). 
Evidently, in the really populated regions we have not more than 10 \% 
difference. 

Comparing this with the much more evident 
differences between the scatter plots 
of protons and antiprotons, 
it suggests that at these energies 
the production of pions is (inside Pythia) 
related with mechanisms that 
conserve scarce memory of the initial state. Else, we should find a 
much  more forward distribution for positive pions (as in the $p/\bar{p}$ 
case).

\begin{quote}
\begin{tabular}{|r|r|r|r|r|r|}
\hline
 Angle & $0^o - 10^o$ & $10^o-20^o$ & $20^o-30^o$ & $30^o-40^o$ & $40^o-50^o$ \\
 Energy & & & & & \\
\hline
0-1 GeV & 12 & -1 & -10 & -7 & 0 \\                 
1-2 GeV & 9 & 0 & 2 & -13 & 0 \\                          
2-3 GeV & -3 & -4 & 3 & & \\                              
3-4 GeV & 1 & -11 & & & \\                                
4-5 GeV & 12 & & & & \\ 
\hline
\end{tabular}
\\
\\Table 4: percentual difference between the bin populations   
of positive and negative pions, in the most populated bins  
(200-2600 charged pions per bin). 
\end{quote}

\section{Effects of the PANDA forward cutoff}

The most relevant weak point in the detection of all the 
fragments of a Drell-Yan measurement is the forward dead ``cone'' 
of the PANDA apparatus. Precisely, it is not a circular-section 
cone, so that in the region of polar angles 
larger than 10 degrees we have full 
acceptance, reduced acceptance (depending on the azimut)
in the region 5-10 degrees, and no 
acceptance at all below 5 degrees. 

Since all the events that could be interesting for theoretical 
developments involve more than one particle, the possibility that 
a particle in the dead cone reduces seriosuly the acceptance 
on multiparticle events is high. As shown below, this is not the 
most relevant problem. 
The biggest problem is the possibility that an event with $N$ particles 
is confused with an event with $N+1$ particles, where one is not seen.

For estimate purposes, the asymmetric forward dead ``cone'' is 
approximated by a true circular cone (i.e. with circular section). 
So I use an average acceptance cutoff at 7.5$^o$ (no 
acceptance at all below this cut, full acceptance over it). 
With this cutoff (see Table 5) we have a drastic reduction of the 
detectable antiproton number, while the other charged particles 
(pions and protons) are less affected by this. 

\begin{quote}
\begin{tabular}{|r|r|r|r|}
\hline
 & All & particles with & cutoff-surviving \\
 &  & $\theta$ $>$ 7.5$^o$ & fraction \\
\hline
$p$ & 25766 & 22540 & 87 \% \\                          
$\bar{p}$ & 25762 & 13370 & 52 \% \\                              
$\pi^\pm$ & 15223 & 12452 & 82 \% \\                                
$\gamma$ & 34696 & 29316 & 84 \% \\                                
\hline
\end{tabular}
\\
\\Table 5: Number of charged particles surviving a forward 
cutoff at 7.5$^o$
\end{quote}

So, I may estimate that 
we will detect 
45 \% of the $p\bar{p}-$only events (0.45 $=$ 0.52*0.87), 
37 \% of the $p\bar{p}+\pi$ 
and of the $p\bar{p}+\gamma$ events 
(0.37 $=$ 0.45*0.83), and so on. 
For events with a not too large number of particles, 
satisfactory rates are not 
a problem, starting from these numbers. The more subtle problem is: 
How much the forward dead 
cone will lead us to confuse events with $N$ and  
with $N+1$ particles? 

Since the 50 \% probability to miss the antiproton regards $all$ 
the below analyzed events with charged particles, 
I will not take this factor into account in most of the 
following\footnote{
The 
probability, say, to confuse an event $p\bar{p}$ 
with an event $p\bar{p}\pi^+\pi^-$ in first approximation 
does not depend on the 
acceptance on $p$ or $\bar{p}$. 
At a finer level of analysis, because of the 
kinematical conservation laws there can be a correlation between 
the probabilities of observing e.g. a proton and a pion in given 
kinematical conditions. This correlation is neglected 
here. 
}. 
In addition, I will assume that those events where $two$ light 
particles enter the dead cone may be neglected. So, 
the noise to ``$p\bar{p}-$only'' events  
is constituted by ``$p\bar{p}+1$'' events, 
the noise to ``$p\bar{p}+1$'' events is constituted  
``$p\bar{p}+2$'' events and so on. 

Last, it is assumed in the following that all the considered events 
are associated to a detected dilepton 
pair.\footnote{
As 
observed in the previous footnote, there may be a correlation 
between the probabilities of detecting the leptons and the 
fragments. This correlation is neglected
here. 
} 

\subsection{(A) $p\bar{p}-$only events.} 

This class contains 10747 events of the simulation, i.e. 
about 21 \% of the total and 
49 \% of all the events where a $p\bar{p}$ 
pair is present. As above stated, I will assume that 
among the remaining 51 \% only those containing $one$ 
extra light particle (charged pion or gamma) 
may easily be confused with events of the 
$p\bar{p}-$only class. 

The events of the class ``$p\bar{p}$ $+$ 1 light particle'' are 2743, 
and in 82-84 \% of the cases the extra light particle is detected. 
This means that in about 466 cases  
we will not see it. 
The noise-to-signal ratio is then 
466/(10747$+$466) i.e. 4 \%. 

\subsection{(B) $p\bar{p}+1$ events. }

The case where the ``1'' is specifically a pion is considered 
later. Here charged pions and photons are equivalently counted. 

The starting point are 2793 ``true'' $p\bar{p}+1$ events, 
of which 
we only see 2327 $=$ 2793 $-$ 466 (see the previous case (A)). 

We have 5428 events with $p\bar{p}$ and two 
light particles. Assuming that 83 \% is the average detection 
probability for a photon or pion, we have 0.83*0.83 $=$
0.69 probability to detect both. So, 31 \% of the above 5428 
events, i.e. 1683 events, 
will show one light particle only. 

So only 42 \% (i.e. 1683/$(1683+2327)$) of the apparent 
number of $p\bar{p}+1$ events   
really belongs to this class. 
 
\subsection{(C) $p\bar{p}+2$ events.}

We have 5428 true $p\bar{p}+2$ events, of which we detect 
5428 $-$ 1683 $=$ 3745. 

Three light particles appear in 1521 events, and the probability 
to detect all three is $0.83^3$. Neglecting the subtraction of 
the 2nd-order 
set overlap (i.e. of the events where $two$ of these three particles 
are not detected) this means a probability 0.43 to see two 
particles only, so 654 apparent $p\bar{p}+2$ events. So the 
fraction of false $p\bar{p}+2$ events is $654/(654+3745)$ $=$ 
15 \%. 

Clearly when we become more specific about the class of 
the two light particles ($\pi^\pm$ or $\gamma$) accompanying 
dilepton and $p\bar{p}$, the contamination 
ratio decreases, but 15 \% is a good upper limit for 
the noise-to-signal ratio.

\subsection{(D) One charged pion (independently 
from the number of associated photons).} 

The previous difficulty in distinguishing events with one 
light particle is much decreased if we exclude photons 
(so also neutral pions) from 
our analysis. Indeed, from table 3 we see that the large 
number of events with two light particles, compared with 
events with one, is due to a large number of photons in the 
two-light-particle case. On the contrary charged pions respect the 
``natural hyerarchy'' since  
charged-pion pairs are 3 times less abundant 
than single pions. 

On the other side, to identify 1-pion-inclusive events is essential 
to be able to indentify $n\bar{p}-$inclusive and $p\bar{n}-$inclusive 
events (see point (F) below). 

We have 7836 events with one charged pion, among which only 
7836*0.82 $=$ 6426 events will be indentified experimentally. 
Then we have 2814 events with two charged pions, among which there is a subset 
$2814*(1-0.82^2)$ $=$ 922 
events that appear as one-pion events (or as no-pion events, I will here 
neglect this less frequent possibility). 
This means a fraction $922/(922+6426)$ $=$ 13 \% of false single pion 
events.

\subsection{(E) $n\bar{n}-$only pairs}

Events where $no$ particles of any kind are 
identified
%\footnote{ 
%Presently the PANDA setup does not include a hadronic shower 
%calorimeter, but unpublished estimates suggest that there is 
%enough matter in the volume inside the magnet to directly 
%identify a fraction $\sim$ 10 \% of the neutrons from the 
%fragments of their 
%strong interactions with the e.m. calorimeter, 
%and much more in the case of segmentation 
%of the external iron layers. 
%Here I do not consider this 
%possibility. 
%}
(apart for 
the dilepton pair) may be identified with $n\bar{n}-$only 
events, in principle. 

These events constitute 20 \% of the total 
(Table 2), 
and they are easy to identify, since the noise is given either 
by events with a single photon, or with a pair of opposite-charge 
hadrons. An estimate like those of the previous points 
suggests that these contaminations are not huge. 

However, these events tell us little, missing more direct 
information. 
It must be observed that the total 4-momentum of the $n\bar{n}$ 
pair is automatically determined since it must 
exactly balance the dilepton 
4-momentum. We cannot however access information 
on the relative configuration of $n$ and $\bar{n}$, so to know 
that an event is of $n\bar{n}-$only nature does not 
give us access to new observables.   

\subsection{(F) Mixed nucleon-antinucleon pairs (inclusive 
identification of the pair)} 

The 7993 events with a ``mixed'' pair 
are events where we see a proton not accompanied by an antiproton, 
or viceversa, and an odd number of charged pions. Here I only 
consider the case of one accompanying charged pion.  

I assume that the numbers of $p\bar{n}$ and $n\bar{p}$ are the same, 
i.e. about 4000. The main error comes from $p\bar{p}+2\pi$ pairs where 
$p$ or $\bar{p}$, $and$ one of the two pions, are not detected. 

Identifying a 
$p\bar{n}$ event on the ground that we find a proton 
unaccompanied by an antiproton (i.e. without checking for the pion) 
would be really unsafe, since 2000 protons coming from a $p\bar{n}$ pair 
are overcome by about
12,400 $p\bar{p}$ events where the antiproton is not seen 
because of the forward dead cone. This means about 0.14 probability 
for identifying a mixed pair from a single proton. 

In the $n\bar{p}$ case half of the true 4000 events is not 
detected because of the antiproton in the dead cone. The left 2000
events must be matched to 3200 $p\bar{p}$ events where the 
proton is not seen. So, when we see an antiproton without a companion 
proton we have 38 \% probability that it comes from an $n\bar{p}$ 
pair. 

So, in both cases we need to identify the accompanying pion  
to have a clear signature of the event, 
hoping that we are not confusing a single-pion event with 
a double-pion event. This confusion, as estimated previuously, 
takes place in 13 \% of  
the single-pion detections, but in this case 
this number can be divided by two. Indeed, 
the pion accompanying the mixed pair must carry a well defined 
charge, while in the case of a missing pion from a pion pair 
the previously calculated 13 \% factor includes 
both charges with equal probability. 

So, to quantify 
the chances that a $\bar{p}\pi^+$ pair 
is associated to an 
$n\bar{p}$ mixed pair we need to compare 
(1-0.13/2)*0.38 to (1-0.38)*(0.13/2). The latter is 
the probability that we have found a $p\bar{p}$ pair accompanied by 
two pions, and is much smaller than the former. So the identification 
of the detected $\bar{p}\pi^+$ pair as an event containing 
an $n\bar{p}$ pair is 
quite safe. 

In the case of a proton and a negative pion, we have to match
(1-0.13/2)*0.14 vs (0.13/2)*(1-0.14), i.e. 0.13 vs 0.06. This means that in 
1/3 of the cases our identification of an event containing a 
$p\bar{n}-$pair from 
a detected $p\pi^-$ pair is wrong. 

It must be observed that in the mixed-pair case the 
accessible information is larger than in the $n\bar{n}$ case, since 
we have some knowledge of the reciprocal orientation of the 
produced fragments. On the other side, these fragments are at least 
three (with possibly additional dead cone photons), 
with potential difficulties in theoretical modelling. 

\section{Conclusions}

The limiting feature of Drell-Yan in PANDA (its reduced energy) 
may be converted into a strength point. We have the possibility 
to analyze in detail more complex and exclusive events where the 
Drell-Yan dilepton pair is accompanied by hadronic/e.m. fragments 
that are recorded in detail. The limited phase space of this experiment 
makes the number of accompanying fragments small. 

An analysis of the involved statistics, after taking PANDA acceptance 
into account, shows that some features of the simplest fragment combinations 
(in particular: of a pure proton-antiproton pair) may be studied with 
reasonable safety: the statistics is reasonable, the danger of 
confusion with events of different nature is small.

%%%%%%%%%%%%%%%%%%%%%%%%%%%%%%%%%%%%%%%%%%%%%%%%%%%%%%%%%%%%%%%%%%%%%%%%%%%%

%%%%%%%%%%%%%%%%%%%%%%%%%%%%%%%%%%%%%%%%%%%%%%%%%%%%%%%%%%%%%%%%%%%%%%%%%%%%%%

\end{document}